\documentclass[twocolumn,twoside,slac_two]{revtex4}
\usepackage{graphicx}
\usepackage{fancyhdr}
\begin{document}

\title{Constraining pulsar gap models with light-curve shapes of a simulated $\gamma$-ray pulsar population}

\author{Marco Pierbattista \& Isabelle A. Grenier}
\affiliation{Laboratoire AIM, CEA-IRFU/CNRS/Universite? Paris Diderot, Service dÕAstrophysique, CEA Saclay, 91191 Gif sur Yvette, France \& Universit\'e Paris Diderot}
\author{Alice K. Harding}
\affiliation{NASA Goddard Space Flight Center, Greenbelt, MD 20771, USA}
\author{Peter L. Gonthier}
\affiliation{Hope College, Department of Physics, MI 49423, USA}

\begin{abstract}
The Fermi Gamma-Ray Space Telescope has discovered many $\gamma$-ray pulsars, both 
as radio-loud objects and radio-quiet or radio-weak pulsars that have been identified through blind
period searches. The latter presumably have $\gamma$-ray beams originating from high altitudes in the 
magnetosphere, resulting in little or no overlap with the radio beam. The exponential cut-off of the
emission at high energy also points to a medium- or high-altitude origin of the $\gamma$ rays. Population 
synthesis offers means to study two complementary aspects of the  $\gamma$-ray pulsar population:
energetics and light curve morphology. The latter (peak multiplicity, position, separation, asymmetries, ...) can be used to constrain the beam(s) geometry (their origin within the open magnetosphere, evolution of the radiating region with pulsar power and age, ...). It can also be used to constrain the obliquity and orientation of a pulsar in the framework of a given accelerator model. We present preliminary results from new simulations and lightcurve analyses that indicate that the sample of Fermi pulsars retains much of the geometrical information of the parent $\gamma$-ray pulsar population.
\end{abstract}

\maketitle

\thispagestyle{fancy}

\section{Introduction}

Rotating, magnetized neutron stars are natural unipolar inductors, generating huge electric fields in vacuum, building a large surface charge, and pulling charges out against the star gravity. Accelerated particles and subsequent photon-pair cascades tend to build a force-free density in the magnetosphere, but particle acceleration requires some local departure (gap) from force-free conditions. Different gaps have been imagined for different assumptions in the global current circulation. Two types of accelerators have been proposed: vacuum gaps above the polar cap and near the light cylinder, and the space-charge limited flows (SCLF) where a voltage develops due to the small charge deficit between the real charge density and the Goldreich-Julian one (\cite{gj69}). Here we study 1) the low-altitude gap that forms in the SCLF a few stellar radii above the polar cap (nicknamed as the \textit{polar cap} (PC) below, \cite{mh03}); 2) the higher slot-like gap that forms in the SCLF along the last closed field line where particles continue to accelerate and radiate to nearly the light cylinder in the unscreened electric field (alias the \textit{slot gap} (SG),  \cite{mh04}); 3) the vacuum \textit{outer gap} (OG) located between the surface of null Goldreich-Julian density and the light cylinder (\cite{crz00} \& 
\cite{yr95}). We do not consider the outer gap version which assumes external currents flowing through the gap region (\cite{h06}). 
It is evident that the narrow hollow cone of radiation expected from the deep regions above the polar cap, the flaring funnel-like beam from the slot gap, and the fan beam from the outer gap will sweep differently across the sky, so pulsar lightcurves potentially provide a wealth of information on the gap location, cascade development and its radiation processes. Yet, flux limited samples of radio and $\gamma$-ray pulsars, and the lack of orientation information, may provide a very biased point of view of the morphological diversity of lightcurves and of the actual radiative power delivered by pulsars in different wavebands in their evolution. Population studies offer a statistical means to test the source location and aperture of the beam(s), as well as the radio and $\gamma$-ray luminosity dependence on pulsar power and age. In these proceedings, we concentrate on morphological properties of the light curves.

\section{Simulation recipes}

We have simulated a population of normal isolated pulsars using a Gaussian distribution for the initial period and a log-normal distribution for the initial magnetic field. The neutron star motions have been evolved in the Galactic potential. Geometrical and luminosity models of the core and cone components of the radio beam have been implemented to compare with the observed properties of the radio pulsars in the ATNF\footnote{http://www.atnf.csiro.au/research/pulsar/psrcat/} catalogue. Details of the radio simulation are given in the accompanying poster by Gonthier et al. (eConf C091122). The standard vacuum dipole spin-down not being able to reproduce the ATNF data, other scenarios have been explored (with a 2.8 Myr timescale for magnetic field decay, or without decay, but correlated initial periods and magnetic fields). Both reproduce the ATNF data reasonably well. We restrict the presentation below to the first case. 

\begin{figure}[!ht]
\centering
\includegraphics[width=0.5\textwidth]{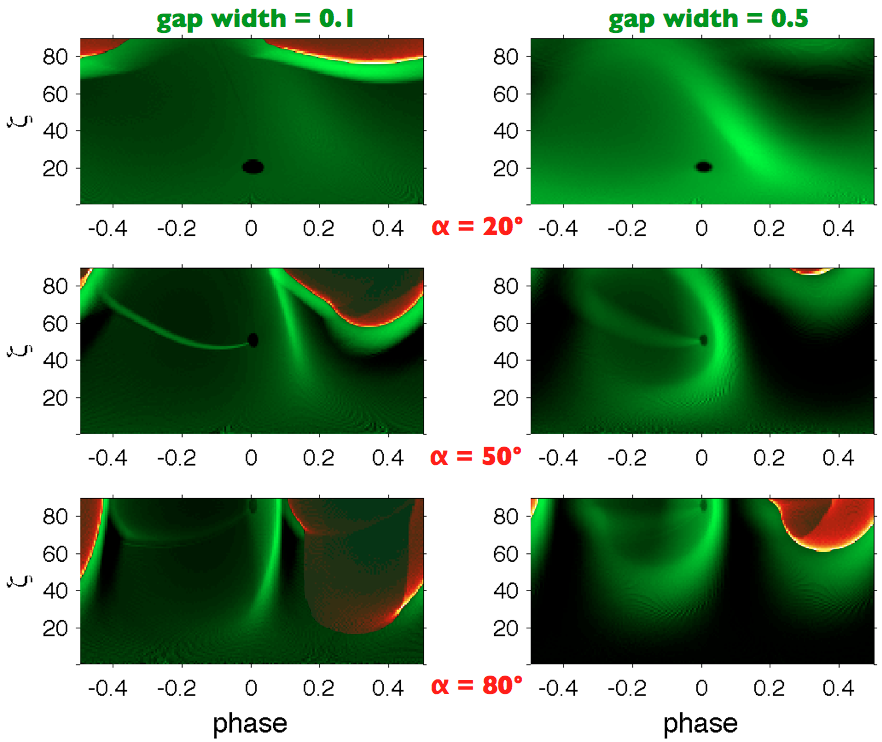}
\caption{  Phas-eplot examples obtained for the SG model (Green) and OG model (red).} \label{PHPLOT}
\end{figure}

In the low-altitude part of the slot gap, pair cascades develop at several stellar radii above the polar cap and produce a wider hollow beam than in the original vacuum-gap polar-cap model because of the flaring of the field lines. The analytical solutions of (\cite{mh03}) for the electric potential and charge density have been used to integrate the particle luminosity across the open magnetosphere. The high-altitude solutions of (\cite{mh04}) have been used to compute the particle luminosity from the higher regions of the slot gap. For both the Ôpolar capÕ and Ôslot gapÕ situations, the total luminosity scales to first order as $L \propto  w_{SG}^3 \dot{E}$ with different scaling factors at low and high altitudes. $\dot{E}$ and $w_{SG}$ denote the pulsar spin-down power and slot-gap width or thickness, respectively. The latter, i.e. the co-latitude at which the pair formation front curves upward, is numerically solved for each pulsar following (\cite{mh04}).

The same luminosity dependance with $\dot{E}$ and gap width applies to the outer gap (\cite{zcjl04}). In this case, the gap thickness is limited by pair production on soft X rays produced by the heated polar cap. It depends on the average altitude of the $\gamma$-ray emission inside the gap (between the null surface and the light cylinder) and on the distance across field lines where thermal X rays and $\gamma$ rays produce pairs (optical thickness of 1). The primary $\gamma$ rays that trigger the pair cascade are assumed to be produced along the last closed field line. 
To produce an atlas of light curves, \cite{wrwj09} have assumed simple relations for the outer-gap high-energy luminosity $L$ and width $w_{EG}$ where $L \propto w_{EG} \dot{E}$ and $w_{EG} \propto (10^{26} W / \dot{E})^{1/2}$. It was motivated by the observations of young $\gamma$-ray pulsars. We have also simulated this variant of the outer gap (nicknamed the \textit{external gap} (EG) hereinafter).

For all models, the gap widths are computed as a function of period, its derivative, and magnetic obliquity. All stars have a moment of inertia $I = 1.8 \, 10^{38}$ kg m$^2$ derived for a 13 km radius and 1.5 solar mass star (\cite{lp07}). They have random viewing orientations and obliquities. A simple radiative efficiency has been used to convert particle power into $\gamma$-ray luminosity (30\%, 100\%, 50\%, and 20\% for the PC, SG, OG, and EG models, respectively, to match the Fermi luminosity constraints). 

\begin{figure}[!ht]
\centering
\includegraphics[width=0.5\textwidth]{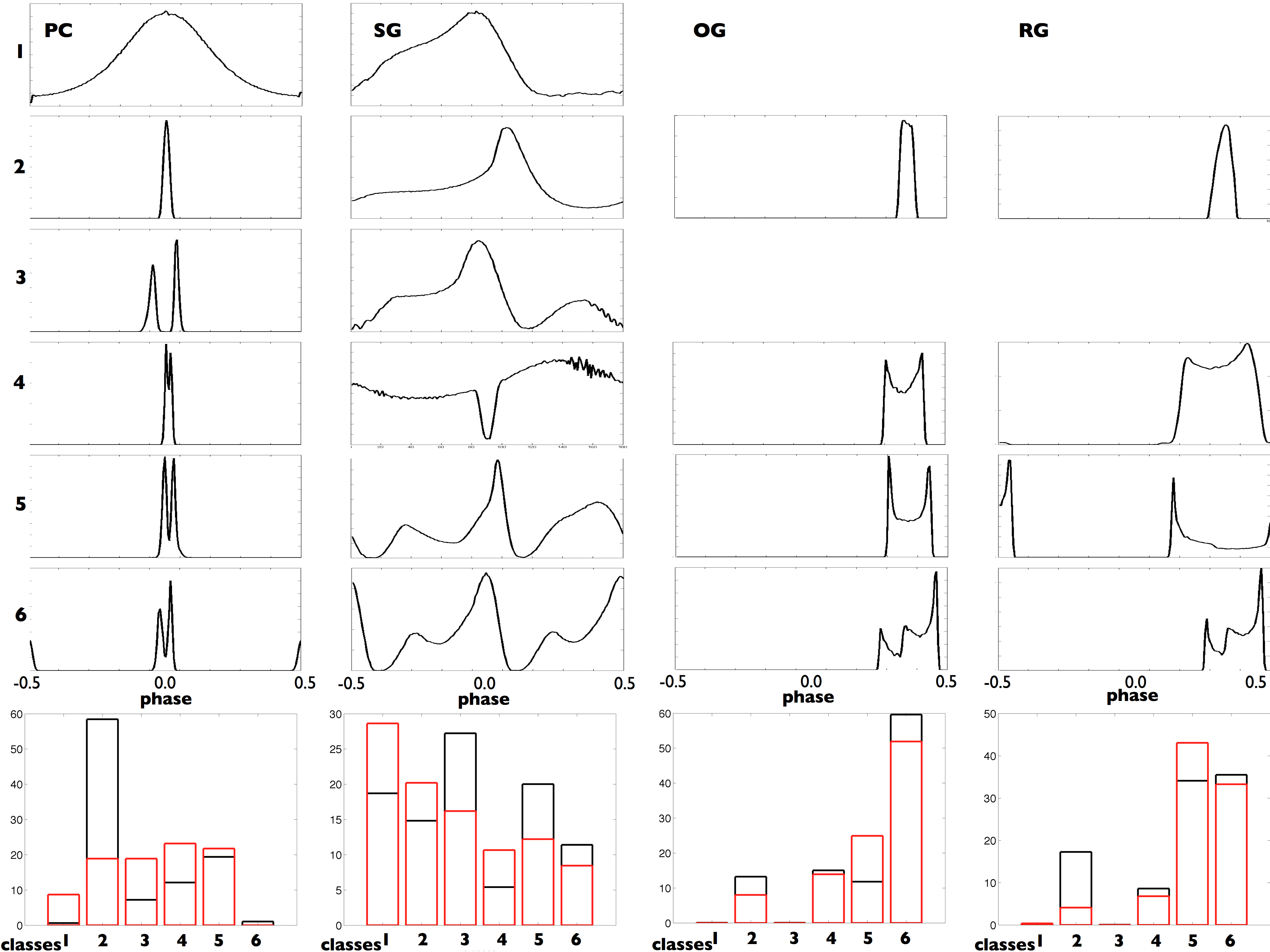}
\caption{ Typical examples of lightcurve shapes found for each model. The histograms give the recurrence of each class in the parent sample that shines towards us (in black) and in the visible $\gamma$-ray sample that passes the Fermi detection threshold (in red).} \label{classes}
\end{figure}

To describe the radiation pattern across the sky resulting from the sweeping of each beam, we have computed 2D phaseplots in the ($\zeta$, $\phi$) plane of the observer viewing angle $\zeta$ and rotational phase $\phi$ (fig. \ref{PHPLOT}). The calculation includes field retardation assuming the vacuum  solution (\cite{d55}), light travel delays, and relativistic aberration (\cite{dhr04}). The sharp peaks seen from the outer gaps profiles in Figure \ref{classes} result from the assumption of a very thin radiative layer along the gap inner edge (away from the last closed field line). In the SG case, we have assumed a broader photon distribution across the gap width. Their density profile perpendicular to the field lines peaks in the middle of the gap and falls to zero along the last closed line and along the gap inner edge. This assumption does not reproduce the extremely thin peaks observed by Fermi in the bright pulsar lightcurves (e.g. Vela or Crab, \cite{aaa+09}), but it reflects the broader radiation pattern that can be expected from wide gaps in older pulsars.
A grid of phaseplots has been produced to fully sample the magnetic obliquities and pulsar characteristics relevant to each radio and $\gamma$-ray beam model (period and its derivative, magnetic field, gap width, radio frequency). The phaseplot of a particular pulsar has been obtained by careful interpolation within the grid, preserving the continuous transformation of pattern shapes from one choice of parameter to the next. The phaseplot has then been normalized to twice the total luminosity per pole that the pulsar can sustain for a given model. Lightcurves, at the viewing angle of the simulated star, have been integrated to compare the average point-source $\gamma$-ray flux with the 6-month sensitivity map given in the Fermi pulsar catalogue ( \cite{aaa+09}). 

\section{Lightcurve shape characterization}
Lightcurve shapes have been analyzed for key characteristics. Average, skewness (asymmetry measure) and kurtosis (peakedness measure) values for a whole lightcurve are easy to compute, but they do not map well the large diversity of shapes. Beyond the kurtosis, we have introduced a sharpness index equal to $log_{10}(100 \times F_{max} / F_{average})$ and we have built an automated classification scheme based on the number of maxima, minima, non-zero bands, and inversion points in the profile. The lightcurves were first smoothed to keep only the bulk features, i.e. a single peak, two well-separated peaks, a closer double peak, triple peaks, shoulders, etcÉ  that are displayed in  Figure  \ref{classes}. The lightcurves have then been fitted with a combination of Gaussians (for the PC) or of Gaussians and Lorentzians (for the SG, OG, and EG) in order to locate the peak positions in phase and to derive their intensity and width. The closest magnetic pole is set at phase zero.

\section{Results and discussion}
Since there are many parameters to explore the lightcurve diversity (peak multiplicity and separation, sharpness, skewness, radio-$\gamma$ peak lags, etc) in this multivariate space ($\dot{E}$, gap width, age, $\alpha$ obliquity, $\zeta$ orientation, etc), we present and comment on only a few snapshots taken from two samples, the total \textit{parent} population of neutron stars with non-zero lightcurve, and the \textit{visible} $\gamma$-ray sample with an integrated point-source flux that is larger than the 6-month Fermi sensitivity threshold in the pulsar direction. The comparison between the parent and visible samples is less sensitive to orientation effects than to flux thresholding since the pulsars from both samples do shine in our direction.

Figure  \ref{classes} gives examples of profile shapes and their frequency in the two samples. 
\begin{figure}[!ht]
\centering
\includegraphics[width=0.5\textwidth]{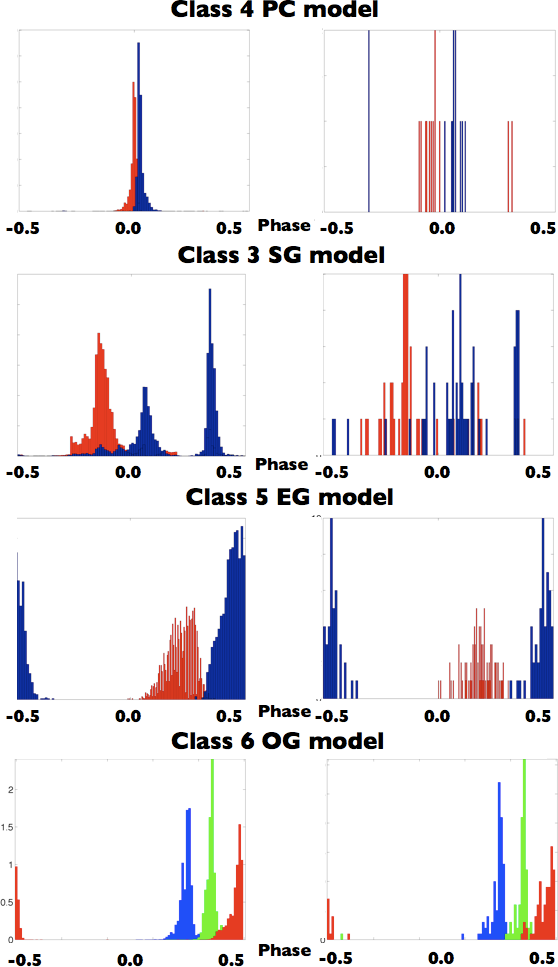}
\caption{ Peak phase histograms of the most recurrent shape classes in each model. In the left plot column are plotted
the phase histograms  of the parent population while in the right one are plotted the phase histograms of the visible
$\gamma$-ray pulsars. For the PC, SG \& EG models the red and blue colors are respectively referred 
to the position of the first and second peak of the light curve. For the OG model is plotted the position of the triple peak.} \label{positionhisto}
\end{figure}
For the Ôpolar capÕ, broad lightcurves with a large duty cycle are rare because they are only  seen when both the line of sight and magnetic axis are close to the spin axis ($\alpha \sim \zeta < 20^{\circ}$). The symmetrical situation ($\alpha \sim \zeta \sim 90^{\circ}$) which gives rise to two peaks separated by half a rotation (class 6) is equally rare. Classes 2 to 5 reflect the fact that the compact, hollow, PC beam is little distorted by delays and aberration when viewed from Earth. Two peaks appear with various degrees of separation as our line of sight crosses closer or further from the beam axis. The single peaks in class 2, seen at angles grazing the beam edge, are fainter, therefore much less frequent in the visible sample than in the parent one.

For both outer gaps, the largely dominant frequency of triple peaks results from seeing the caustic effects along both the leading and trailing field lines. The change in relative frequency between class 5 (double peak + low bridge) and class 6 (triple peaks) in the OG and EG gaps is due to different evolutions of the gap thickness with age. The middle peak appears at large magnetic obliquities and strengthens in intensity with increasing gap width (older age). The OG sample being richer in thick-gap objects than the EG sample, triple peaked  lightcurves are more frequent.

Few triple peaked  lightcurves have, however, been found among the Fermi pulsars ( \cite{aaa+09}). This discrepancy may be due to the low $\gamma$-ray statistics that prevent the detection of a fainter third peak, yet the first and middle peaks often have comparable intensities in class 6. More exciting alternatives imply that the leading-side gap(responsible for the middle peak) is somehow less efficient than the trailing gap, or that the emission intensity decreases at very high altitudes near the light cylinder. So, the presence of a third peak in the data may become a useful test of the outer gap energetics, provided that photons are produced in a thin layer compared to the gap width. Thicker emission layers would smear the middle peak out (\cite{vhg09}). 

Slot-gap emission fills the whole sky, thus all phases in a lightcurve. Broad class 1 lightcurves are frequent because they can arise for $\alpha < 30^{\circ}$ and $\zeta > 80^{\circ}$ for thin gaps, as well as at $\zeta < 50^{\circ}$ and all $\alpha$ for thick gaps. Sharper single peak (class 2) lightcurves appear at small to medium obliquities for small viewing angles, but only for thin enough gaps from young pulsars. The other, more complex profiles imply crossing first the superposed radiation borne at high altitude on the trailing side of one pole and the leading side of the second pole, then the caustic on the trailing side of the second pole. Double class 6 features appear at large $\alpha$. So, the complex radiation pattern in the two-pole context of the SG yields a large variety of light-curves which are distinct from the outer gap or polar cap ones, but with a rather flat distribution in frequency that cannot be easily used to find orientation or obliquity trends among the Fermi pulsars.

\begin{figure}[!ht]
\includegraphics[width=0.5\textwidth]{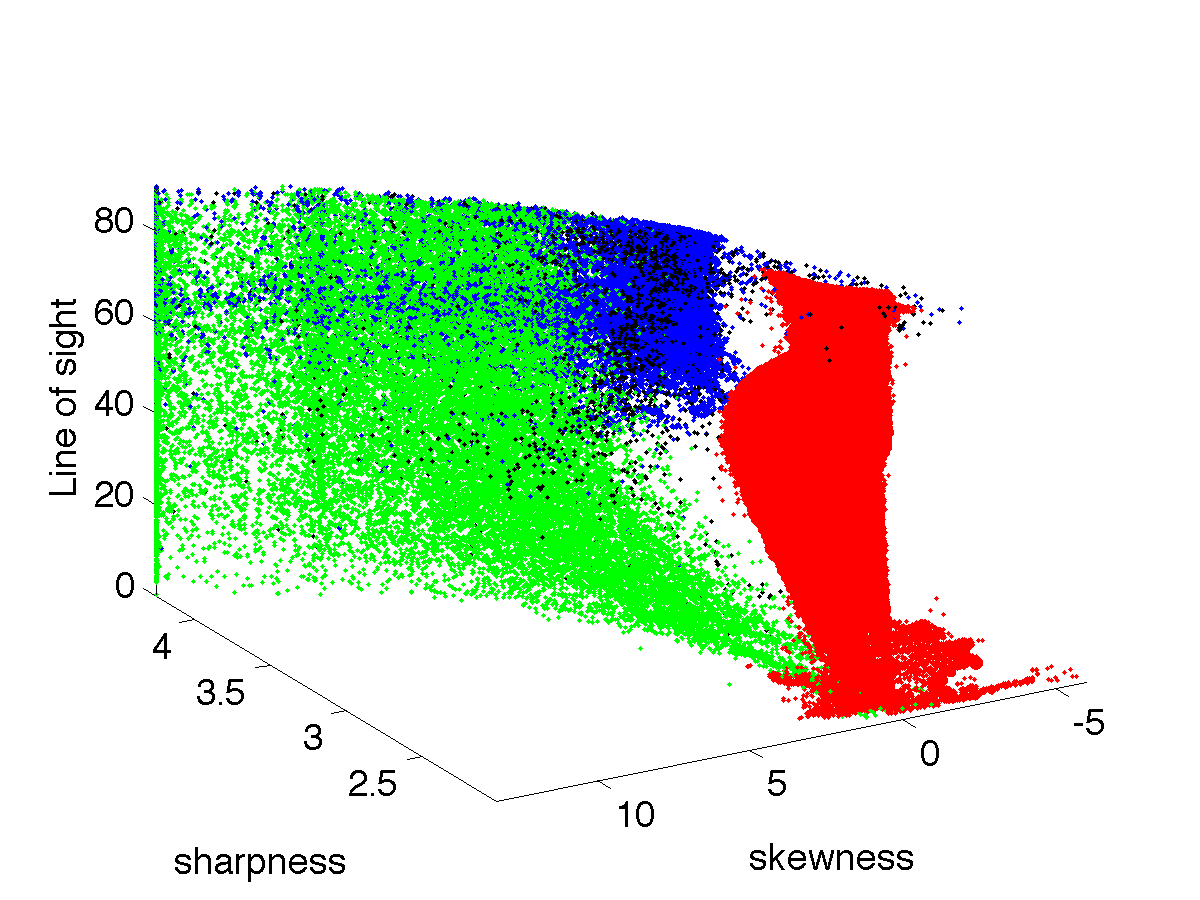}
\includegraphics[width=0.5\textwidth]{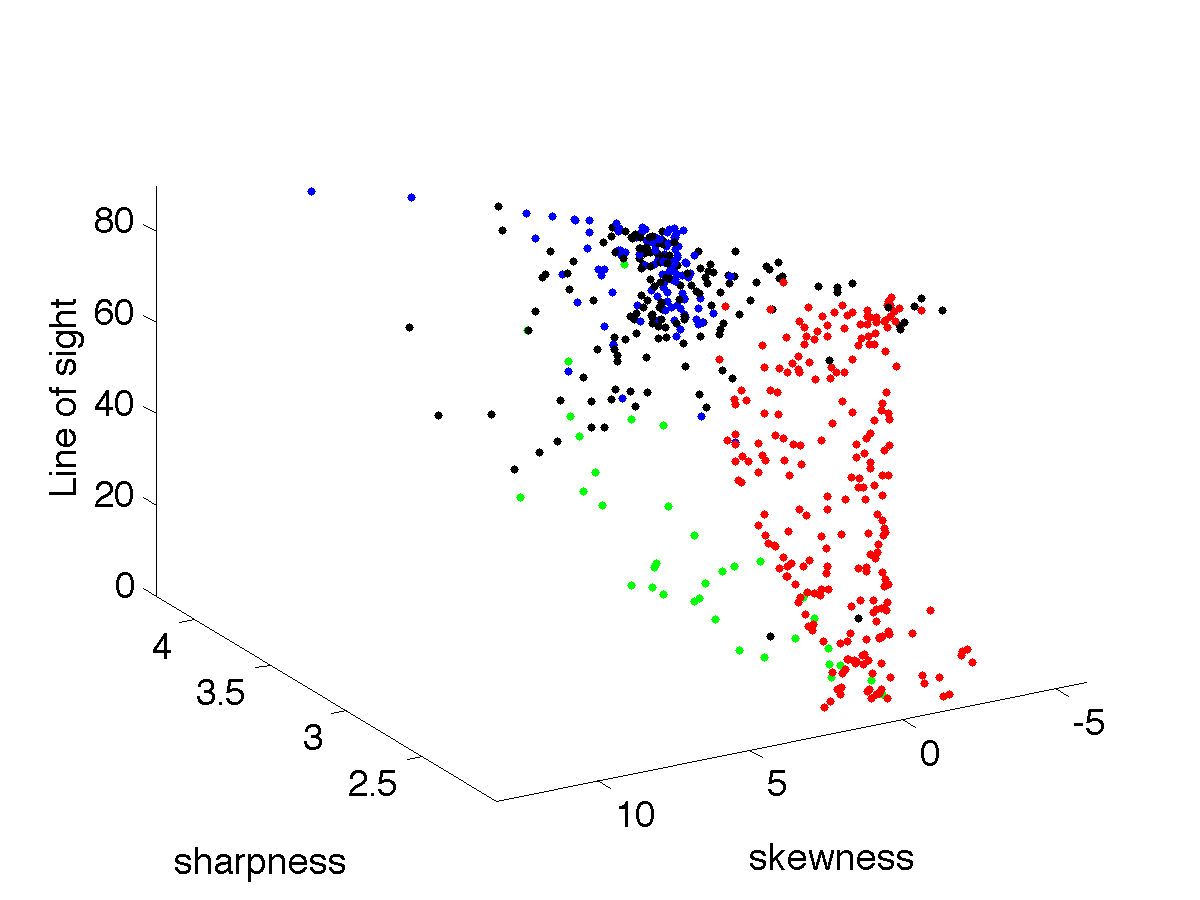}
\caption{Pulsar distribution in the 3D sharpness, skewness, and $\zeta$ viewing angle phase space. The sharpness index is defined as the percentage of the maximum peak intensity to the lightcurve average, in log scale. The green, red, blue, and black represent the PC, SG, OG, and EG models, respectively. The parent and visible samples are shown respectively in the first and second plot.} \label{3Da}
\end{figure}

The series of plots in Figure \ref{positionhisto} shows the histograms of the peak positions in phase for a representative class for each model (class 4, 3, 6, and 5 for the PC, SG, EG, and OG models, respectively). The plots show that the visible sample nicely retains much of the peak position information present in the parent population, thus also the peak separation and lags information. The other classes behave similarly. Admittedly, the Fermi pulsar sample may be biased to sharply peaked and/or small duty cycle light-curves because their periodicity is easier to detect. This effect is not taken into account in our visible sample. We do not, however, expect it to strongly modify our results since most of the Fermi pulsars that have been identified from their modulation have also been found as un-pulsed point sources and that the faintest ones have integrated fluxes in good agreement with the sensitivity threshold (see Figure 11 in \cite{aaa+09}). In other words, we have not yet found many strongly modulated pulsars below the point-source sensitivity threshold that would hamper the comparison with our simulated visible sample. 

Figure \ref{3Da} illustrates our hope to use lightcurve morphology to help discriminate between models when we have estimates of the viewing angles. The latter can be derived, for instance, from pulsar wind torii images in X rays. In the $\zeta$, skewness, and sharpness index phase space, the cloud of black EG points (sometimes coincident with blue OG ones) extends in front of a curved sheet of blue OG and green PC points. The SG red points little overlap with the others. The sharpest PC and OG lightcurves are not retained in the visible sample because their small duty cycle yields a faint integrated flux, below the sensitivity threshold of Fermi. The shift to high $\zeta$ from the parent to visible OG sample corresponds to the shrinking of the radiation pattern with gap widening with age. The Fermi lightcurves are being analyzed to compare with this distribution. 

\begin{figure}[!ht]
\includegraphics[width=0.5\textwidth]{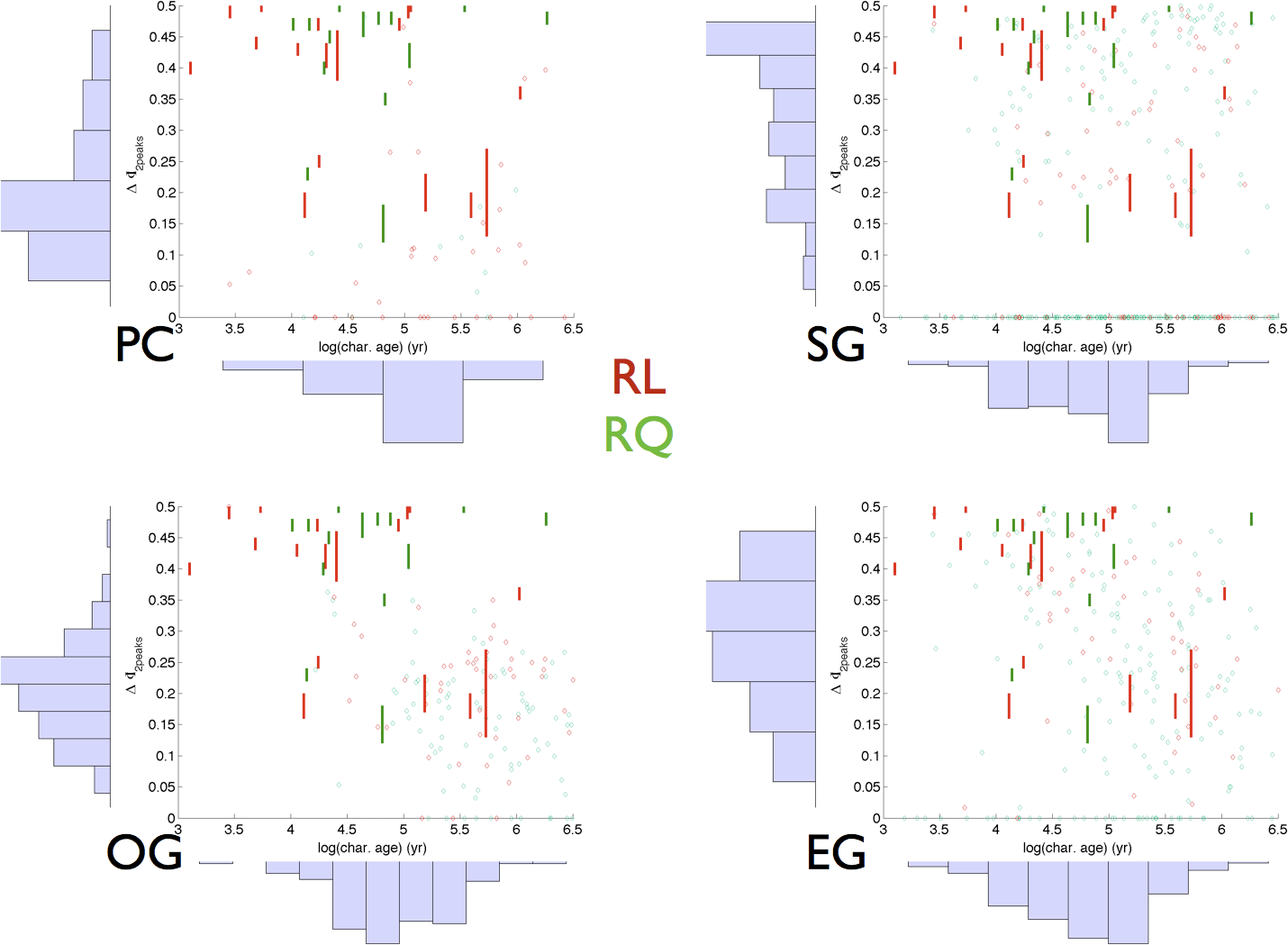}
\caption{ Comparison of the phase separation measured between the two brightest peaks in the Fermi pulsar lightcurves (error bars) and in the 
simulated lightcurves of the visible $\gamma$-ray sample (radio quiet or weak pulsars in green and radio loud ones in red).} \label{2pksep}  
\end{figure}

Figure \ref{2pksep} shows the evolution with age of the phase separation between the two brightest peaks found in the simulated lightcurve, for each model. The measurements made on the Fermi lightcurves are also plotted (\cite{aaa+09}). Three quarters of the observed Fermi profiles were indeed found to exhibit two peaks. The narrow peak separation expected from the PC case is clearly at odds with the data. As the whole outer-gap radiation pattern shrinks with age as the gap inner edge moves away from the last closed field lines, so does the peak separation because both peaks come from a single magnetic pole. Only very young objects with thin outer gaps that dig deep into the magnetosphere can yield widely separated peaks. This is due to the choice of a lower limit of the gap extent at the null surface in the standard OG model geometry. Comparing the OG and EG plots, one can clearly see the impact of the different schemes in gap widening with age. The SG model, which is based on emission from the two poles, favours mostly large peak separations and it does not show much evolution of the peak separation with age. The contrast in photon density within the SG radiation pattern decreases with gap thickness and age, but the bright features due to overlapping emission from the two poles and to the caustics hardly shift in phase. The gap locus along the field lines does not evolve as in the OG gap. The stable regions of photon accumulation in the phaseplot, anchored to the dipole lines configuration, appear to be quite consistent with the Fermi data. So, the peak separation may turn out to be a useful test of the evolution of the gap altitude with age, or of the origin of the visible emission over two poles or a single pole.

\section{Conclusion}
Fermi has at last opened a large bay window on $\gamma$-ray pulsars and the young pulsars cooperate with apparently high efficiencies at converting their spin-down power into $\gamma$ rays that allow us to probe their magnetospheric currents. We have simulated a population of young spin-powered radio and $\gamma$-ray pulsars and computed their lightcurves and flux in the framework of four different high-energy models. The comparison between the population that shines in our direction and the ÔvisibleÕ $\gamma$-ray population that meets the flux requirement for detection by Fermi interestingly suggests that the Fermi pulsars retain much of the morphological lightcurve characteristics that can help discriminate between gap locations and their evolution over $10^{5-6}$ years. 

\bigskip 

\bigskip 










\end{document}